\documentclass[a4paper,11pt]{article}
\usepackage{jinstpub} 
\usepackage{lineno}
\usepackage{subcaption} 

\title{\boldmath Design of an atmospheric muon tomographer for material identification based on CORSIKA+GEANT4 simulations}

\author{Rengifo, J. A}
\author{and Bazo, J. L}
\affiliation{Secci\'on F\'isica, Departamento de Ciencias, Pontificia Universidad Cat\'olica del Per\'u, Av. Universitaria 1801, Lima, Perú}

\emailAdd{jrengifo@pucp.edu.pe}

\abstract{In recent years, muon tomography has turned into a powerful and innovative technique for non-invasive imaging of large and small structures with applications in different areas like geology, archaeology, security, etc. We present the design and simulation of a transportable and easy to construct detector based on plastic scintillator and Silicon photomultipliers current technology. From a flux of cosmic rays reaching the atmosphere we simulated atmospheric muons at ground using CORSIKA. The detector and the object to analyze are simulated with GEANT4, where the previously obtained muon flux is transported. We use two methods for muon tomography to differentiate objects made of different materials: absorption and scattering. The statistical differences for several object sizes and materials are quantified. Using a threshold of 3 $\sigma$ in the first method, we conclude that materials made of lead can be differentiated from objects made of other materials. The observation time needed to differentiate an object made of lead from one of aluminum was 4.9 and 9.9 days using the first and second method, respectively. In general, the absorption method gives the best results.}


\begin{document}
\maketitle
\flushbottom

\section{Introduction}
\label{sec:intro}

Muons, that arrive at ground, created in the atmosphere after the collision of cosmic rays with high energy are highly penetrating, continuously available and non-invasive. These muons can provide information, through their energy loss, absorption and scattering, about the density (i.e. composition) of the traversed materials, thus they are a perfect resource for research. 

In the last 70 years muons have been used to image the inner parts of large or small inaccessible structures as a non intrusive method. For instance, the image of a tunnel of the underground railway in London was accomplished by E.P. George in 1956. He used nuclear emulsions as a detector \cite{George1956}. Later, L.W. Alvarez in 1960 \cite{Alvarez1970} used scintillator counters to image the structure of the second Pyramid of Giza in Cairo, Egypt, trying to find secret passages or hidden chambers. Recently, there have been other studies, as the discovery of a void in Khufu's Pyramid in Giza, Egypt \cite{Morishima}, which used three types of detectors: photographic films called nuclear emulsions, plastic scintillators and gas detectors. Other geology projects, as MU-RAY \cite{Ambrosino_2014}, used three planes of plastic scintillators forming a muon telescope to study European volcanoes, as well as in Colombia \cite{Pena-Rodriguez:2020} and Japan \cite{Ola_2019}. In Japan, the internal structure of Mt. Asama volcano was studied using a cosmic-ray muon detection system and muon tomography \cite{Tanaka}. addition, in security it is possible to look for nuclear material using plastic scintillators with wavelength shifting fiber arrays \cite{Woo2013} and creating images of possible nuclear threats \cite{Morris2014}.

Another study \cite{JungHyun} focused on an image reconstruction algorithm, which use the scattering angle and momentum of muons. This technique could be used for spent nuclear fuel and nuclear material management. Recent works on muon tomography include the design of a muon detector using triangular plastic scintillators with wavelength-shifting (WLS) fiber readouts \cite{Wang_2024} and methods to detect dark matter using muon detectors and application in others areas like environmental sciences, archaeology and civil engineering \cite{Yu:2024spj}.

In this work we explore muon tomography techniques to find differences between blocks made of just one of these materials: aluminum, lead, iron, concrete, water and air. In order to achieve this goal, we design a muon track detector choosing the best geometrical parameters, and simulate an easy to transport mechanical structure (i.e. with small dimensions: 0.48 x 0.48 x 0.2 m$^3$). This detector is made of plastic scintillator and Silicon Photo-Multipliers (SiPMs) current technology. The simulation is based on CORSIKA, for the atmospheric cascades simulation, and GEANT4, for photon detection at the plastic scintillator sensitive area. The absorption and scattering methods are discussed looking for the best statistical significance in material differentiation. Similar works (\cite{Chaiwongkhot},\cite{Bajou},\cite{Barnes}) have been developed; however, this is focused on the capabilities of differentiating between materials, with an angular resolution of 1 $^{\circ}$.

This paper is divided as follows: in Sec. \ref{sec:MuTomo} we describe the muon tomography technique and both methods used for differentiating materials. In Sec. \ref{sec:DesignDetect} the characteristics of the detector are presented. In Sec. \ref{sec:DSimunDetect} both simulation stages: CORSIKA and GEANT4 are described. In Sec. \ref{sec:Results} the results on the significance of both methods for differentiating materials are presented. Finally, in Sec. \ref{sec:Conclu} we give our conclusions. 

\section{Muon tomography}
\label{sec:MuTomo}

Cosmic rays arriving at Earth interact with the molecules of the atmosphere producing particle showers. Muons are the most abundant and energetic of these secondary particles that reach ground and can be easily measured. They arrive with an average energy of 3 to 4 GeV at a rate of 167 particles m$^{-2}$s$^{-1}$ \cite{HAL_Muons}. At these energies, muons can pass through many metres of rock, concrete or other materials. When muons interact with different materials they are attenuated or deviated, properties which are applied in the tomography technique \cite{Lorenzo_2020}, reconstructing the characteristics of the traversed structures in a non invasive way. These methods are explained next.

\subsection{Absorption method}\label{SecAbs}

Muons have a probability to be absorbed by a target along the line of sight, depending on the attenuation of the radiation and the density of the matter traversed. The effective absorption or transmission gives the ratio between the muon flux after passing an object and looking at the free sky. In Table \ref{Table_absor} we show the density for the materials used in this work and the length 1 GeV muons can traverse inside them. 

\begin{table}[ht]
\centering
  \begin{tabular}{||c c c c c c ||} 
  \hline
  Material & Atomic & Atomic & Density &  Traversed length &\\ 
    &  number (Z) & weight (A) & [$g/cm^3$] &  length [$cm$] &\\ 
 \hline
 Air & 7.35 & 14.72 & 1.205 x $10^{-3}$ &  421 x $10^3$&\\
  \hline
 Water & 10 & 18 & 1.000 & 473&\\
 \hline
  Polystyrene & 5.6 & 104.15 & 1.060  & 456.6 &\\
 \hline
 Concrete & 50.53 & 100.53 & 2.300 &  237.4 &\\
  \hline
  Aluminum (Al)  & 13 & 26.98  & 2.699 & 213.8 &\\
  \hline
  Iron (Fe)  & 26 & 55.845 &7.874  & 81.3 &\\ 
  \hline
  Lead (Pb)  & 82 & 207.2 &	11.35  & 72.0 &\\

 \hline
\end{tabular}
\caption{Material properties density and traversed length for 1 GeV muons \cite{GroomPDG2}.}
    \label{Table_absor}
\end{table}

Some projects are based on this techniques, for instance MU-RAY \cite{Ambrosino_2014} studied European volcanoes using three planes of plastic scintillator forming a muon telescope and the Maya Muon project used portable muon detectors to map the interiors of sealed Mayan pyramids in Belize \cite{Fredrick2016}.


\subsection{Scattering method}\label{SecSAD}

When a muon passes through matter its trajectory is deflected by a target nucleus by many small-angle, in most cases, Coulomb scatterings described by the Rutherford cross section. The global scattering angle is the sum of smaller deflections. Using the central limit theorem the distribution of these angles can be approximated by a Gaussian distribution with standard deviation \cite{Lynch_1991}:

\begin{equation}
    \theta_{MCS}=\frac{13.6 MeV}{\beta c p} \sqrt{\frac{X}{X_0}} \left( 1+ 0.038 \ln \frac{X}{X_0} \right)
    \label{rms}
\end{equation}

where $p$, $\beta c$ are the momentum and speed of the incident particle, $X$ is the distance traveled in the material and $X_0$ is the radiation length defined by \cite{Lynch_1991}: 

\begin{equation}
    {X_0} = 716.4 g/cm^2 \frac{A}{Z(Z+1)} \left(\ln\left(\frac{287}{\sqrt{Z}}\right)\right)^{-1}
    \label{X0}
\end{equation}

where $Z$ is the target charge number and $A$ is the target mass number.

If we want to accurately study an object using muon tomography, we will need at least two detectors: one before/above and another after/below the studied structure. The upper detector obtains the initial atmospheric muons, while the lower detector, detects the muons after they have interacted with the structure. However, there are some experiments that use three or four detector planes in order to improve the efficiency and resolution and reach three-dimensional images \cite{Kaiser2019}. 


\section{Detector Design}
\label{sec:DesignDetect}

\begin{figure}
\centering
\begin{subfigure}[b]{0.35\textwidth}
\includegraphics*[width=\textwidth]{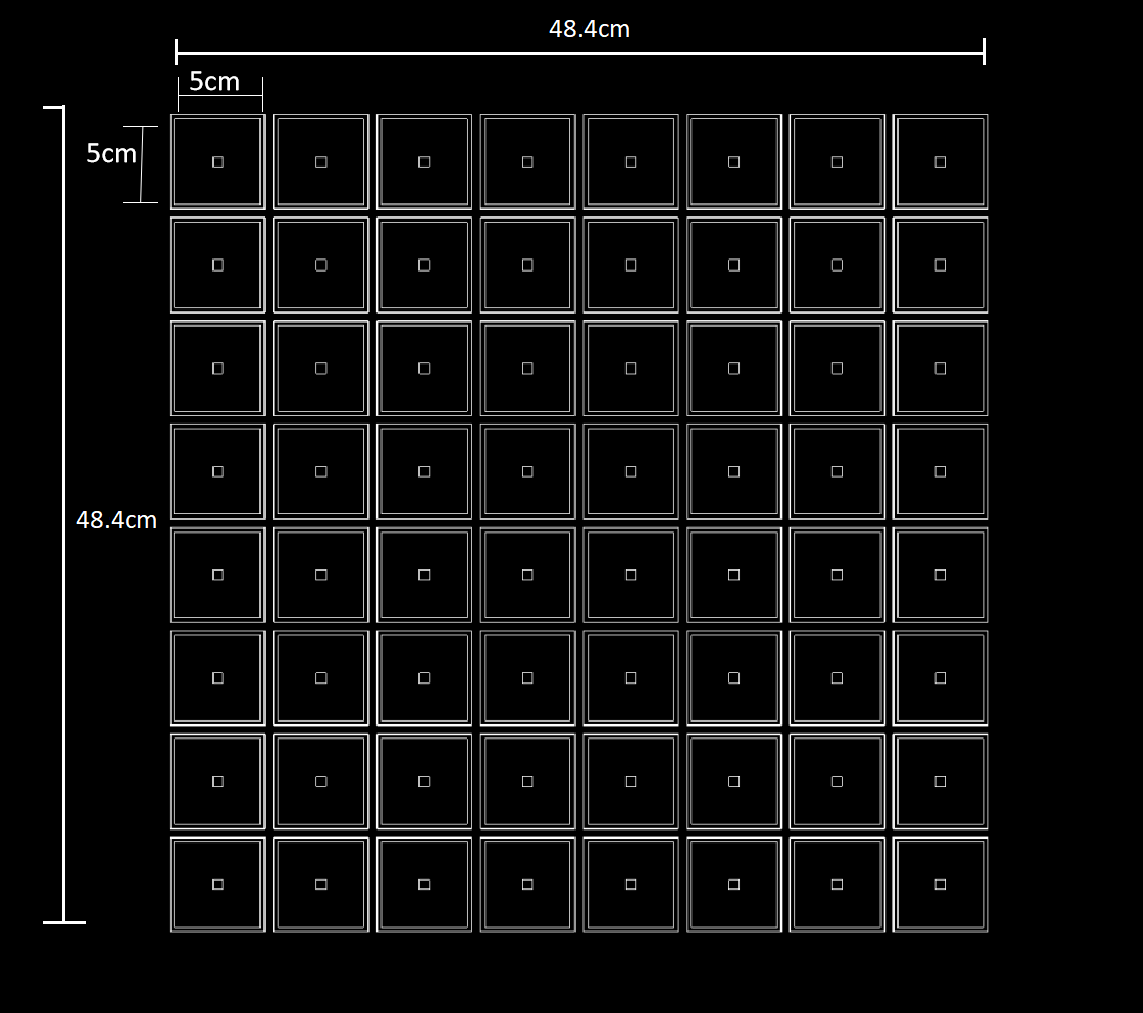}
  \caption{Plane array of unit-sensors.}
  \label{Fig_Sub-Detec}
\end{subfigure}
\hfill
\begin{subfigure}[b]{0.6\textwidth}
\includegraphics*[width=\textwidth]{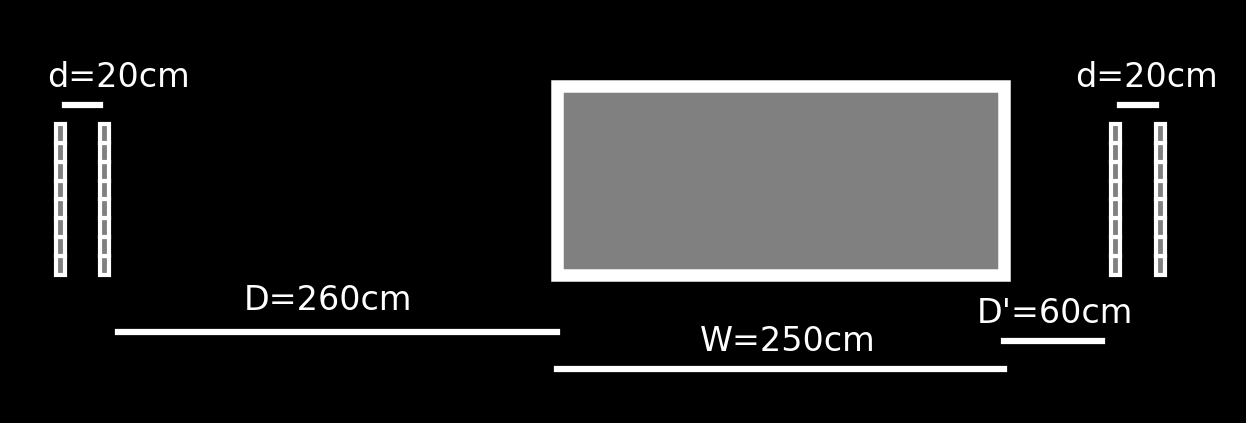}
  \caption{Subdetectors made of two planes and studied object, showing $W$ (object's width), $d$ (distance between two planes), $D'$ (distance between first sub-detector and object) and $D$ ( distance between last sub-detector and object).}
  \label{Fig_Detec}
\end{subfigure}
\caption{Detector design and measurement geometry in GEANT4.}
\label{fig:detector}
\end{figure}

The present detector takes as reference for the basic detection unit the Desktop Muon Detector \cite{Axani:2018qzs, Axani2017}. This unit-sensor consists of an organic plastic scintillator BC408 made of polystyrene ($[C_6H_5CH CH_2]_n$) of 5x5x1 $cm^3$ coupled to a central light sensor (e.g. Silicon photomultiplier (SiPM) of $6\times6$ mm$^2$). As an alternative, a more cost affordable light sensor could be based on commercial CMOS webcams \cite{Hachaj2023}. Since we want a good angular resolution and larger detection area, we design computationally an squared array of unit-sensors making a plane made of 8x8 unit-sensors enclosed in aluminum boxes of 0.05 cm width in order to isolate them, with a total volume of 48.4 x 48.4 x 1 $cm^3$ (see Fig. \ref{Fig_Sub-Detec}) . For a basic reconstruction of the muon track's direction we use two of these planes, forming a sub-detector. To reconstruct the change in trajectory of the muons (i.e. scattering angle) the whole tomographer will be made of two sub-detectors, placed before and after the object under study (see Fig. \ref{Fig_Detec}). Given its relative small size this detector could be easily moved between different locations.

To obtain the best resolution and material identification we searched for the following appropriate geometrical parameters of the measurement setup: the distance between the two planes of each sub-detector $d=20$ cm, the distance between the first plane of a sub-detector and the object $D'=60$ cm and the distance between the last plane of sub-detector and the object $D=280$ cm. This configuration gives an angular resolution of 1$^\circ$ between the object and the detector. We keep these parameters fixed and analyze different widths $W$ and materials of the studied structure, a rectangular parallelepiped of a fixed 80 cm $\times$ 80 cm cross section with a varying width $W$ between 50 cm to 250 cm in 50 cm steps and made entirely of one of these different materials (i.e. lead (Pb), iron (Fe), aluminum (Al), concrete, water and air), to find the size range where materials could be distinguished. This is a simplification of a building, mountain, archaeological remain or smaller objects, for instance, spent fuel casks, shipping containers or nuclear waste containers (\cite{Poulson},\cite{Clarkson}). This approach has limitations since the real objects are more complex, for example, they can have structures with holes, materials combinations and different geometries.

In addition, in order to increase the angular coverage of muon tracks, in our simulation, the sub-detector located after the object could turn 90 $^{\circ}$ approximately around it, which could be implement in a real case using a rotating mechanism remotely controlled. This would allow one sub-detector to be positioned on one side of the analyzed object and then rotated the other one around. The mechanical system for precise rotations would have motors, gears, and a robust framework that supports the sub-detector movement. Before the scanning process, the detector would be calibrated to characterize the response of the scintillator and SiPMs inside the unit-sensors. In addition, we would need alignment verification using muons themselves to guarantee precise data acquisition. 

\section{Simulation}
\label{sec:DSimunDetect}

We simulate cosmic rays that arrive at Earth with CORSIKA and the detector with GEANT4. In this section we will explain the features of these simulations.

\subsection{CORSIKA simulation}
\label{sec:CORSIKA}

We use CORSIKA v76900 (COsmic Ray SImulation for KAscade) \cite{Heck_CORSIKA} for simulating extensive air showers started by cosmic rays. We use for the low energy model GHEISHA (Gamma Hadron Electron Interaction SHower code) and for the high energies QGSJET (Quark Gluon String model with JETs).

We have considered Lima, Peru (12$^o$ 2' 35" S and 77$^o$ 1' 41" W) as the geographical observation site with the following parameters: the observation level was 161 m.a.s.l. and the Earth's magnetic field was 24,8 $\mu T$ for the vertical component ($B_z$) and -0,4 $\mu T$ for the horizontal component($B_x$), from \cite{MF_NOAA}. We used the whole angular range covering zenith from 0${^\circ}$ to 90${^\circ}$ and azimuth from -180${^\circ}$ to 180${^\circ}$. The output of these simulations is the input to the GEANT4 detector simulation. 

We gave to CORSIKA the input cosmic rays fluxes using libraries from \cite{Asorey_Data_Access}, developed by LAGO project, which included the type of nuclei, their spectral indexes for a power law flux and normalization for each nuclei. We simulated the primary cosmic ray flux for one hour, considering the area of the subdetector plane for an energy range from $10^2$ to $10^6$ GeV. This energy range was chosen for a more efficient simulation. Primary particles that arrive at the atmosphere with lower energies after interacting with molecules produce secondaries that cannot arrive at the surface. In addition, primaries with higher energies than the upper limit, arrive at Earth with a very low flux. Finally, for low energies we used the rigidity cut-off 12.17 GV \cite{Rigiditty2021}, as a parameter in CORSIKA. The total number of simulated showers for one hour measurement was 23822. 

\begin{figure}
\begin{center}
    \includegraphics*[width=0.7\textwidth]{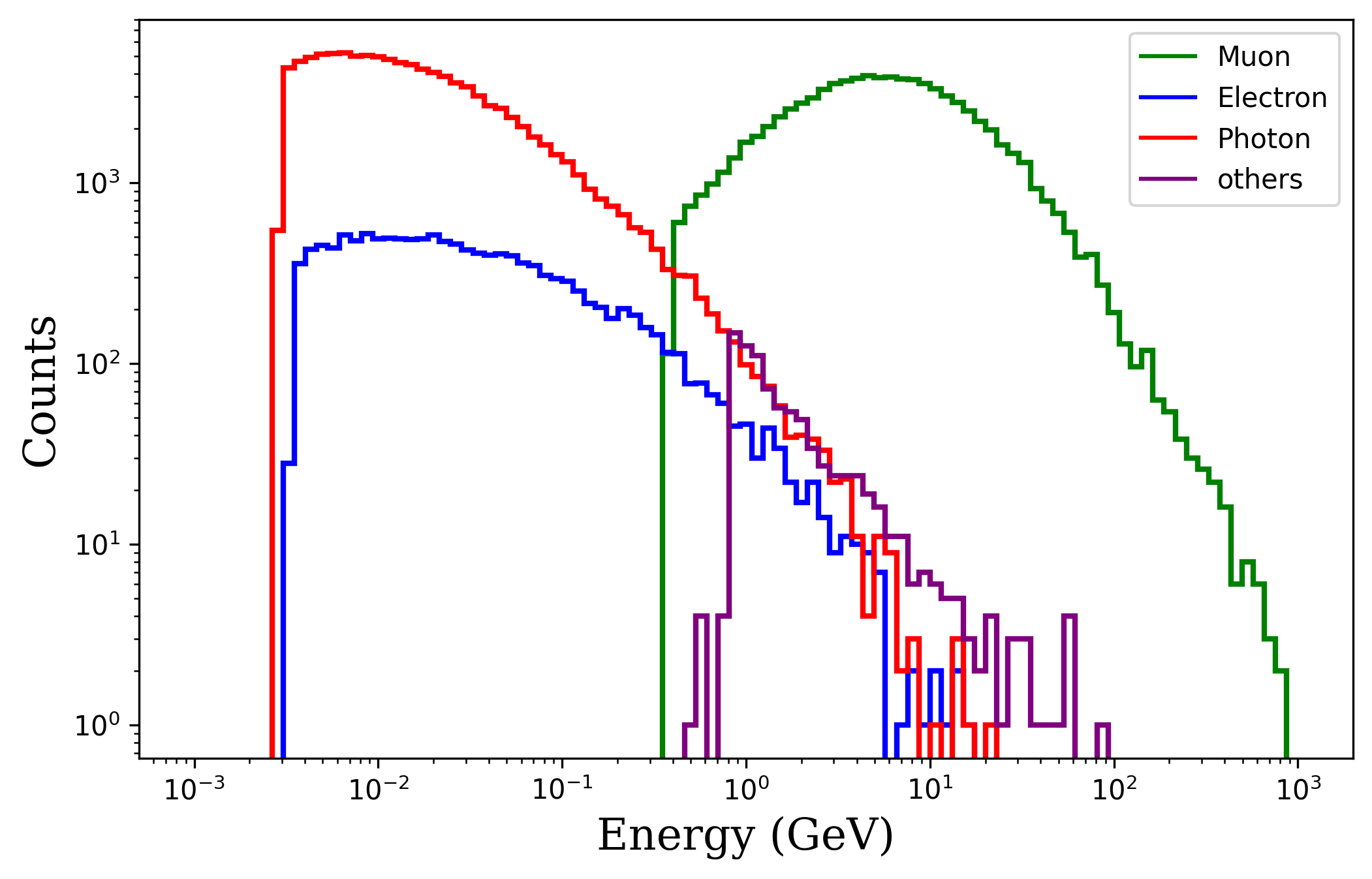}\\
\end{center}
  \caption{Energy distribution of secondary particles arriving at ground generated by the input primary cosmic ray composition.}
  \label{EmvsE_RadSPF}
\end{figure}

We show the energy distribution of particles arriving at ground generated with CORSIKA by the primary cosmic ray composition in Fig. \ref{EmvsE_RadSPF}. Photons are most abundant at MeVs energy as well as electrons, while muons are predominant at GeV energies. The mean energy of muons is 4 GeV. 

\subsection{GEANT4 simulation}

We use GEANT4 (GEometry ANd Tracking) version 10.05 \cite{Agostinelli2003} to simulate secondary particles from cosmic rays passing through the different materials of the experimental setup. The input data is the flux of particles obtained from CORSIKA (i.e. type, position and momentum of the particles generated in each shower). Since muons are more energetic and arrive in larger numbers than other particles at ground, we have chosen to include only atmospheric muons in our simulations.
   
To obtain a larger number of useful events without increasing the CORSIKA production time, we made a modification to the original particle direction. Instead of considering particles arriving from all directions, we have set one direction for all muons aligned with the line connecting the detectors and the object. This modification of the direction of the flux of particles from CORSIKA, also will modify the real observation time, which will be larger than the simulated one. This point is addressed later. 

We defined the experimental setup in GEANT4 by introducing the geometry, materials and sensitive regions, as described in Sec. \ref{sec:DesignDetect} and also define the necessary physics processes. Observed events are those that generate scintillation photons in the active material. 

\begin{figure}[h]
\begin{center}
\includegraphics*[width=0.7\textwidth]{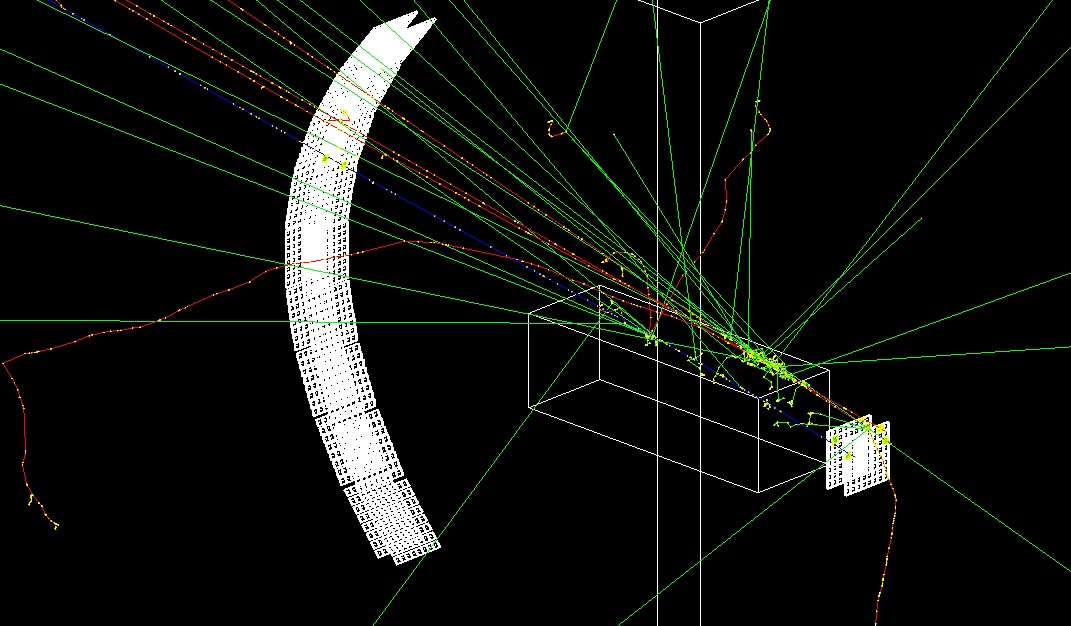}
\end{center}
  \caption{GEANT4 measurement setup showing 10 particles coming from one shower with the second sub-detector turning $\approx90^{\circ}$ around the central point. Red lines are negative charged particles (e.g. muon tracks and electrons), blue lines are positive particles (e.g. antimuons tracks and positrons), while green lines are photons.
  }
  \label{Fig_DesignDMD}
\end{figure}

In order to simulate the sub-detector rotation, we have replicated the sub-detector copying it four times below the original one and four times above it, forming an arc. In Fig. \ref{Fig_DesignDMD} we see the geometry of how muons pass through the detector. When they hit the block, some are absorbed, some are diverted and some reach the sub-detector that will rotate to obtain multiples angles for larger statistics.

On the other hand, we would need a time-tagging system to find a correlation of hits in our detectors to reconstruct the trajectory of muons. In the simulation we do not consider any systematic uncertainties due to reconstruction efficiency related to a time stamp. We use the MC truth (i.e.  particle number) for each event to reconstruct its direction given the unit detector where the energy was deposited.

\section{Results}
\label{sec:Results}

We have analyzed the following two muon tomography methods for material differentiation using simulations:
\begin{enumerate}
   \item Absorption method: fraction between $N_i$ (particles that pass through the first sub-detector and should have arrived to the last sub-detector, verified through their direction in air) and $N_f$ (particles that actually arrive after passing the block).
   \item Scattering method: width of the Gaussian fit to the scattering angle distribution. 
\end{enumerate}

To quantify the statistical difference between measurements of different materials we calculate the number of sigma deviations ($\sigma$) between them: 

\begin{equation}
\sigma = \frac{\| x_A-x_B \|}{\sqrt{(Er_{x_A})^2 + (Er_{x_B})^2}}
    \label{NS}
\end{equation}

where $x_A$ and $x_B$ are the variables for each different material (i.e. A and B) and $Er({x_A})$ and $Er({x_B})$ are their corresponding errors. This variables and their errors in the case of the first method are the fractions of the absorption method and in the second method are the width of the Gaussian fit to the scattering angle distribution.

\subsection{Absorption method}

\begin{figure}    
\centering
\includegraphics*[width=0.7\textwidth]{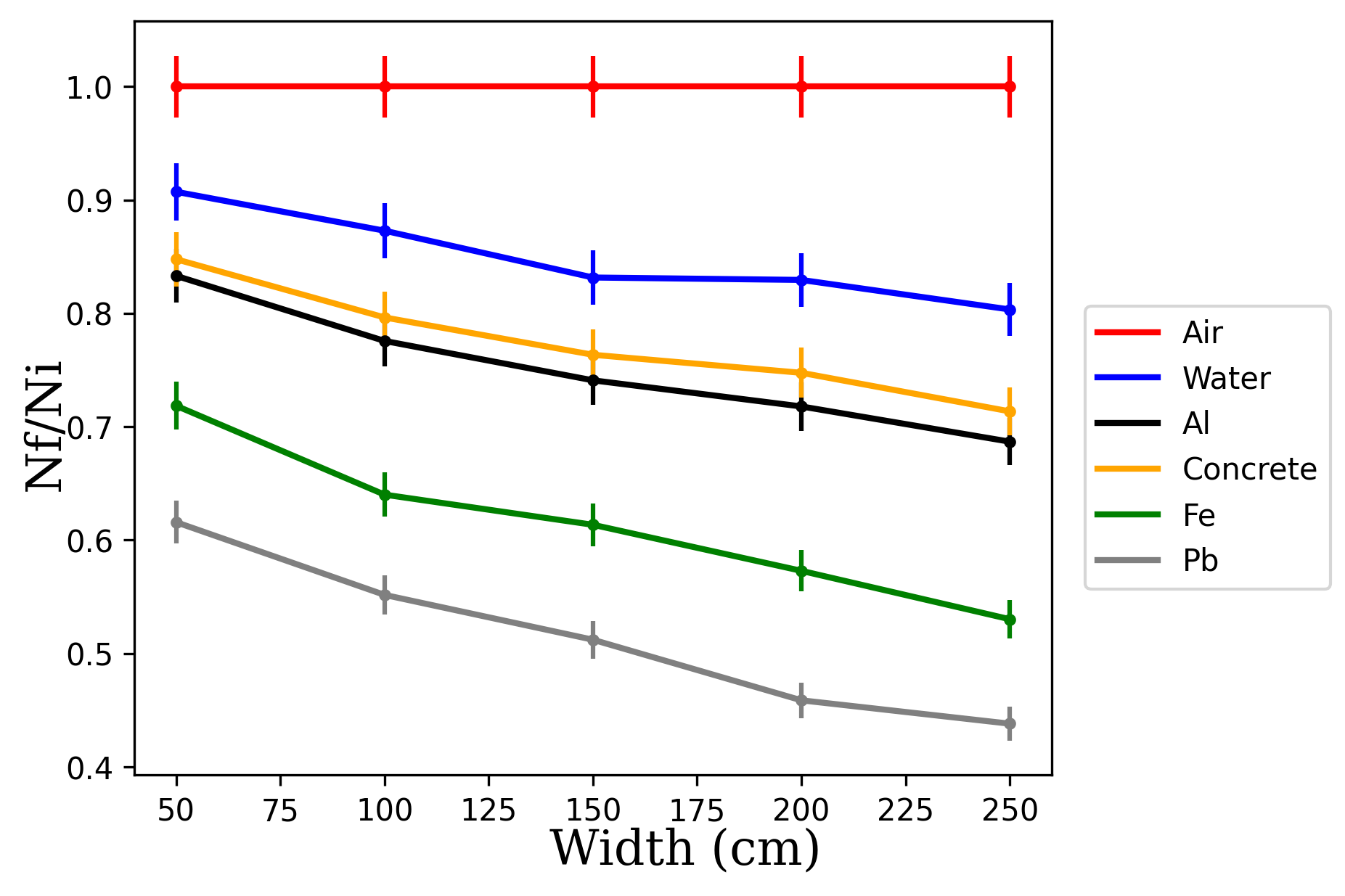}
   \caption{Absorption Method: fraction of particles detected in the sub-detector after the object and those that should have arrived after passing through air for different materials as a function of the object width.}
    \label{Fig_Fraction}
\end{figure}

In Fig.\ref{Fig_Fraction} we show the fraction of detected particles $N_I/N_i$ for the absorption method, for several materials as a function of the width of the test block. These materials can be differentiated in most cases, except for aluminum and concrete. Since the air block is the reference its fraction is 1. Then this fraction decreases as the atomic number and density increases. As the block width increases, the fraction decreases, as expected, and the differences between materials becomes more pronounced. 

\begin{figure}
\begin{center}
\includegraphics*[width=1\textwidth]{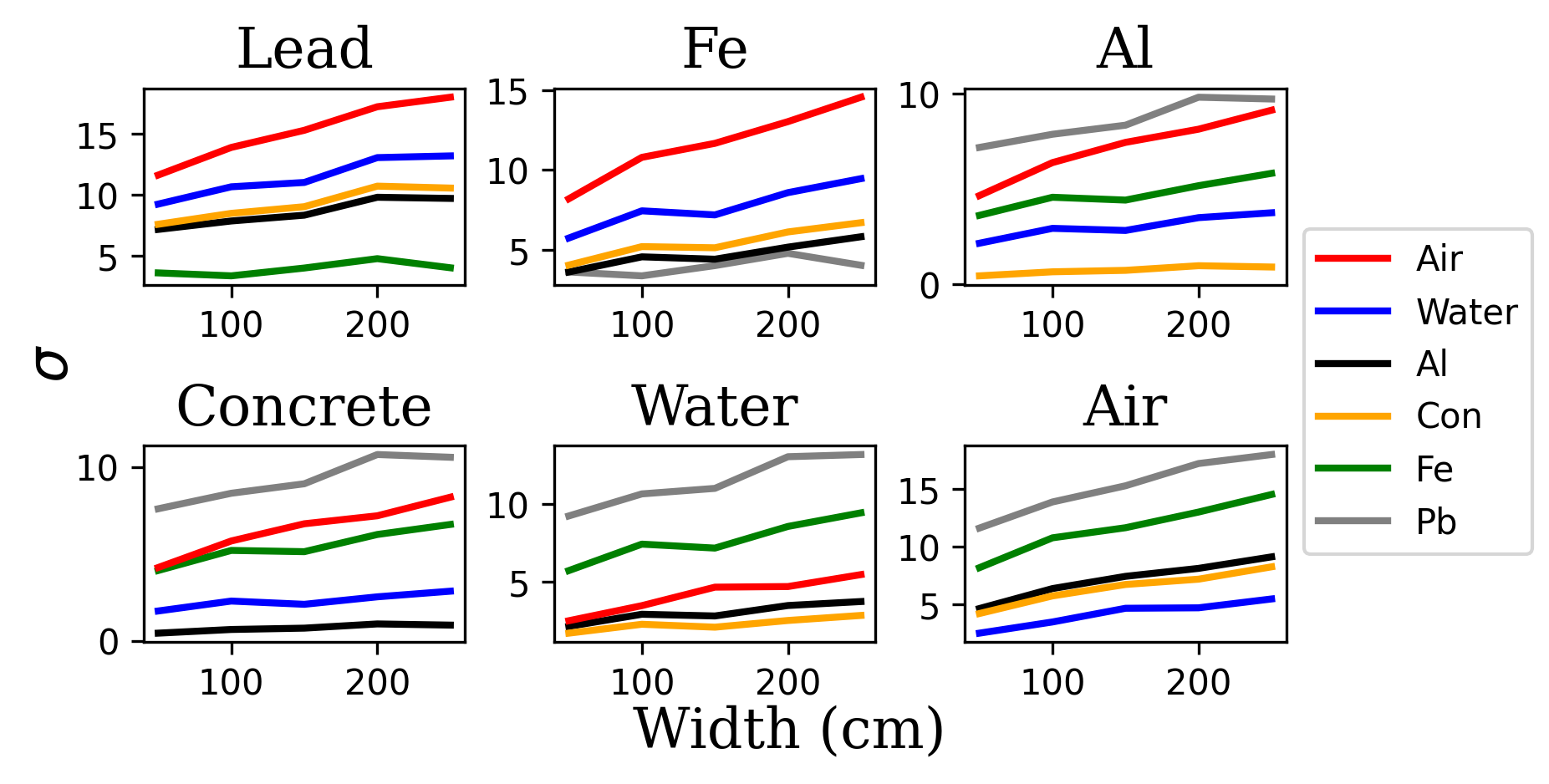}
\end{center}
  \caption{Number of sigma deviations between the material stated on top of each plot compared with those in the legend.}
  \label{Fig_SigmaFraction}
\end{figure}

In Fig. \ref{Fig_SigmaFraction} we present the number of sigma differences between fraction of diverse materials. As an example, the differences between lead and iron ranged from 3.5 to 4.1 $\sigma$, but between lead and aluminum were much more considerable, from 7 to 9.9 $\sigma$. As expected, given their atomic numbers and densities lead is more similar to iron but more different than aluminum. Moreover, the results show a relatively small difference of 0.7 to 0.9 $\sigma$ between aluminum and concrete, indicating that they are similar in terms of material properties. 

\begin{figure}
\begin{center}
\includegraphics*[width=0.7\textwidth]{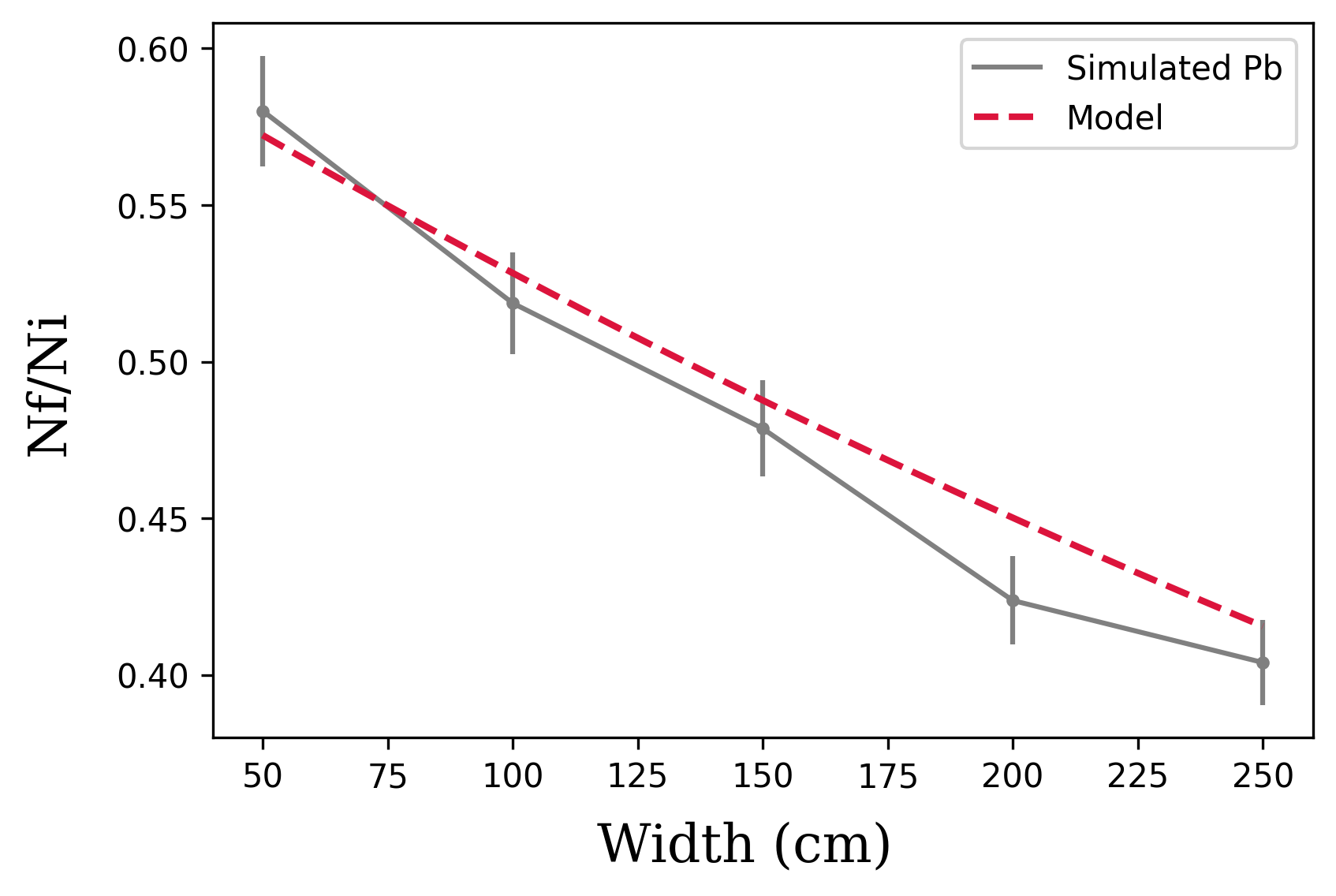}
\end{center}
  \caption{Absorption method fraction comparing the results from our simulation in gray with the theoretical exponential function model in dashed read, for lead.}
  \label{FracW_Theo}
\end{figure}

As an example, we compare the results for lead with the theoretical exponential function in Fig. \ref{FracW_Theo} using the absorption coefficient for muon impacting on lead \cite{ALTAMEEMI2019281}. Both results are compatible within the errors. 

\subsection{Scattering method}

\begin{figure}
\begin{center}
\includegraphics*[width=0.7\textwidth]{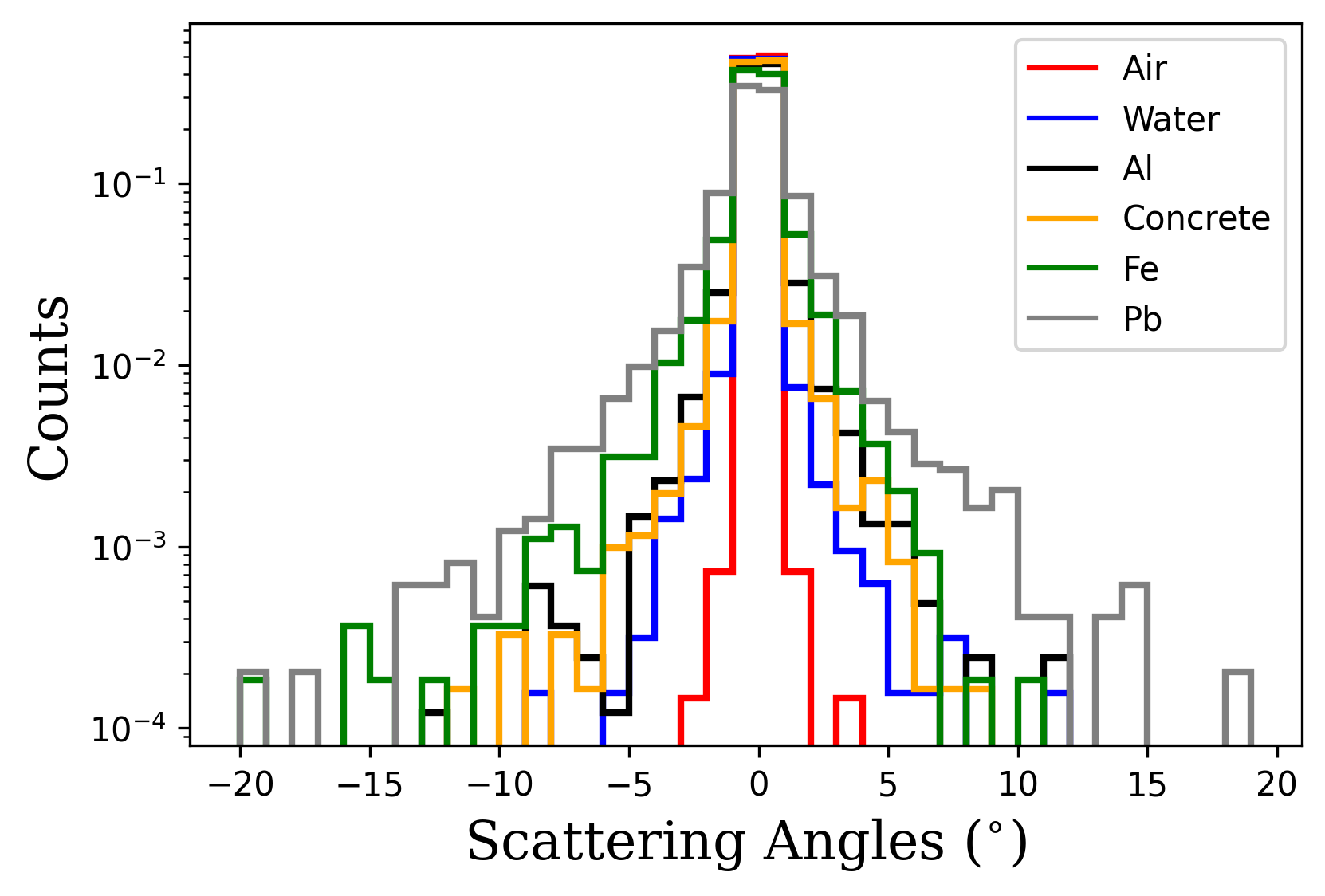}
\end{center}
  \caption{Normalized scattering angle distributions for all materials using a 250 cm block.}
  \label{Fig_DistributionAngles}
\end{figure}

The scattering method evaluates the full width at half maximum (FWHM)of the Gaussian fit to the muon scattering angles distribution. For this calculation we identify the muon hit positions by the emitted scintillation photons in each of the two planes of a sub-detector and calculate the corresponding vector that joins the points of each sensor that is traversed. Using a vector from each sub-detector before and after the block we calculate the scattering angle of the muon. The resulting angular distribution is fitted with a Gaussian function. For instance, for a 250 cm width block, we show in Fig. \ref{Fig_DistributionAngles} the scattering angle distribution for different materials. Lead exhibits a wider distribution compared to iron while for concrete it is much thinner. This means significantly narrower angle dispersion.

\begin{figure}
\begin{center}
\includegraphics*[width=0.7\textwidth]{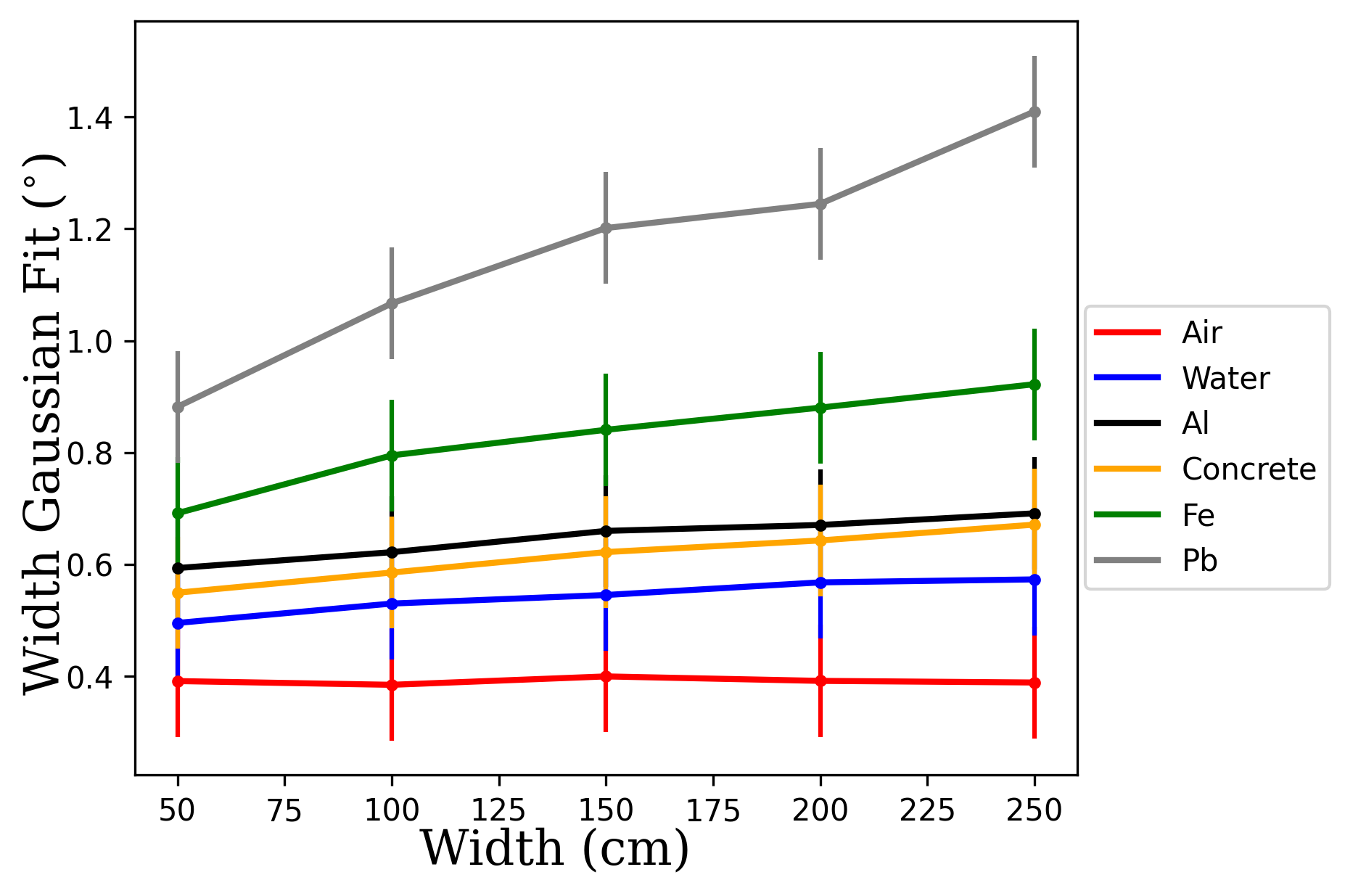}
\end{center}
  \caption{Width of the Gaussian fit of the scattering angle distribution varying the width of the block for different materials.}
  \label{Fig_WidthGauss}
\end{figure}

In Fig \ref{Fig_WidthGauss} we see the influence of the block width on the FWHM. Wider blocks resulted in broader scattering distributions. The lead curve moves away from all materials as the block gets wider, followed by iron. However, other materials are harder to differentiate since their errors superimpose.

\begin{figure}
\begin{center}
\includegraphics*[width=1\textwidth]{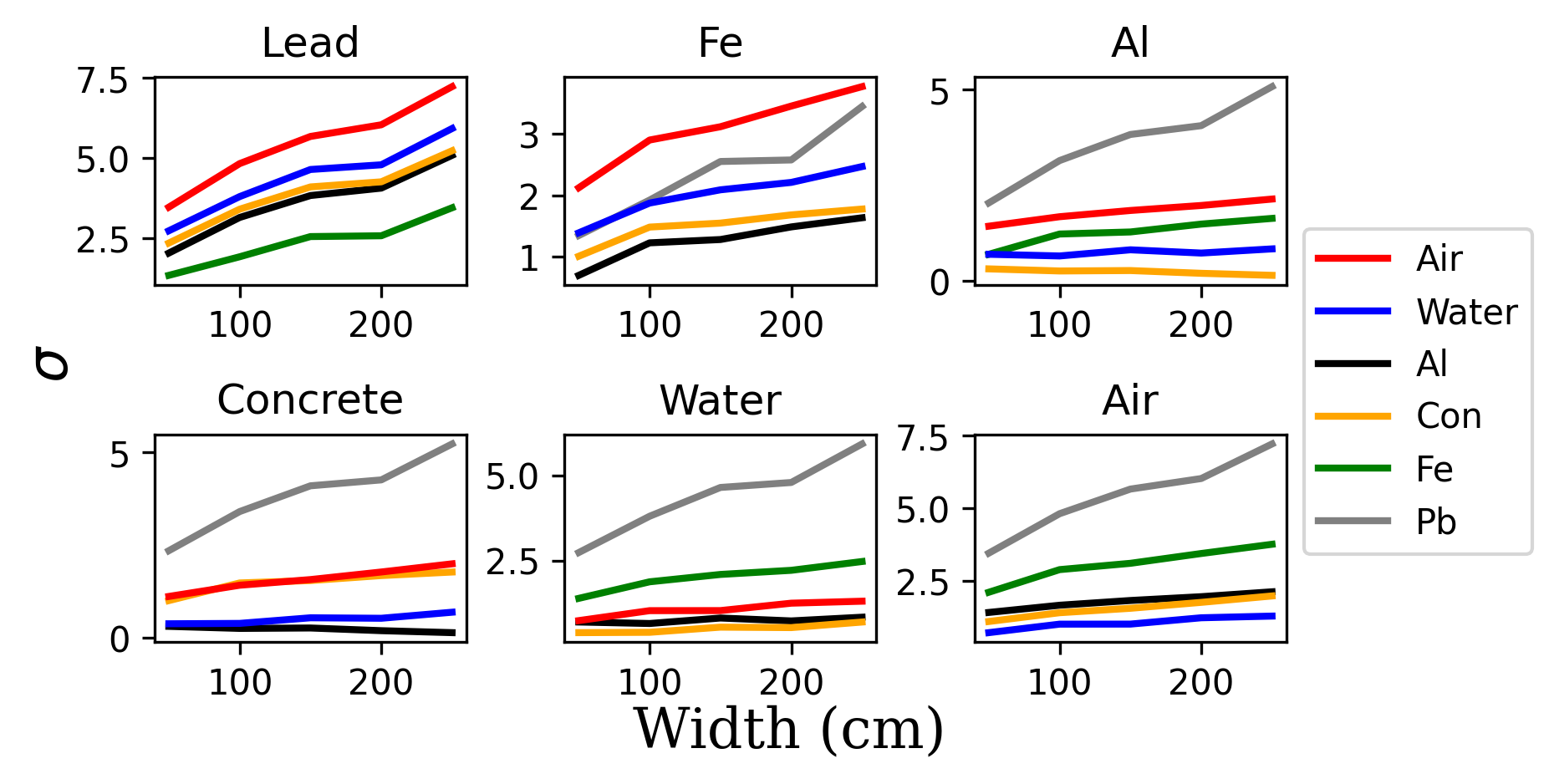}
\end{center}
  \caption{Number of sigma deviation between FWHM of different materials, as a function of the test block width.}
  \label{Fig_SigmaWidth}
\end{figure}

To quantify the difference between the FWHM for each material, we show in Fig. \ref{Fig_SigmaWidth} the numbers of sigma (see Eq. \ref{NS}). For the smallest size of the block (i.e. 50 cm) muons do not deviate much showing a lower number of sigmas. However, as the block width increases, so does the number of sigmas. Because lead is denser and can absorb more muons than other materials, there is a larger number of sigmas between it and aluminum, concrete, water, and air. However, there are very low number of sigmas between concrete and aluminum or water and air. Thus it will be difficult to find differences between these materials, unless there is a larger observation time.

\begin{figure}
\begin{center}
\includegraphics*[width=0.7\textwidth]{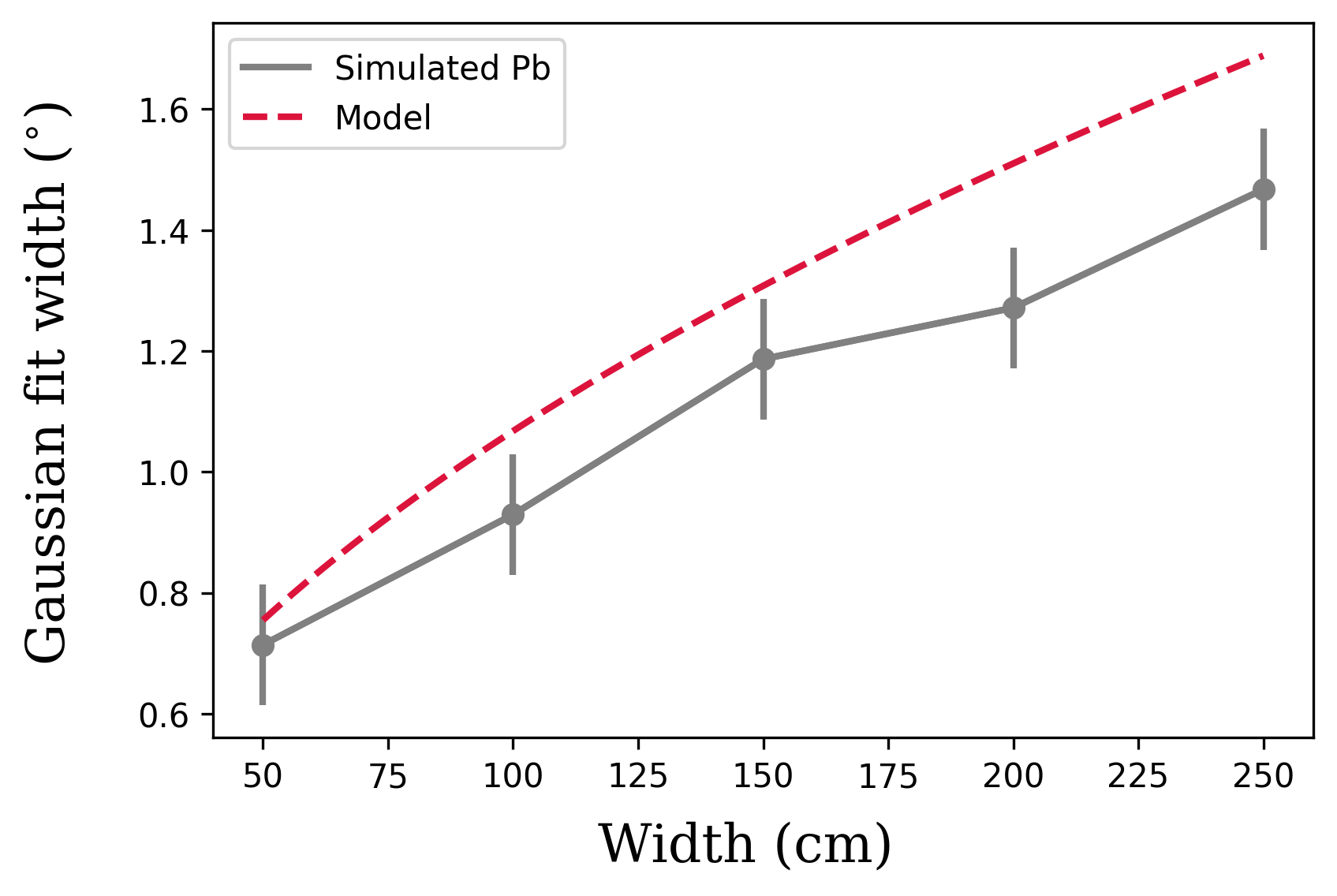}
\end{center}
  \caption{FWHM of lead, as a function of test block size comparing our simulation with the theoretical curve.}
  \label{SigW_Theo}
\end{figure}

We compare for lead in Fig. \ref{SigW_Theo} the simulated FWHM as a function of the size of the block with the theoretical curve (see Eq. \ref{rms}). Both follow the same trend, however, the simulation predicts smaller angles.  

To compare both methods we set 3 $\sigma$ as a common threshold to identify the materials. With the absorption method it is possible do the following differentiation: 
   \begin{itemize}
     \item Between lead or iron and all other materials and widths of the object.
     \item Between air and concrete or aluminium, for all object widths.
     \item Between concrete and water for an object of 200 to 250 cm width.
     \item Between air and water for an object of 100 to 250 cm width.
     \end{itemize}

For the scattering method it is possible to differentiate in the following cases:
   \begin{itemize}
     \item Between air or water for an object of 100 to 250 cm width.
     \item Between air and iron for an object of 150 to 250 cm width.
     \item Between lead and iron for an object of 250 cm width.
     \item Between lead and aluminum or concrete for sizes between 100 to 250 cm.
     \end{itemize}

To quantify the actual measurement time to obtain 3 $\sigma$ we need to consider the following. We simulated one hour of cosmic ray showers in CORSIKA, however to increase the statistics in GEANT4 we included two modifications. The first is the movement of the second sub-detector around the object, which multiplies by 9 the real time, because of all positions are simulated at the same time. The second modification was the initial directions of the secondary cosmic rays which were changed to a single direction. In one hour without the modified directions, 125 muons reached the last sub-detector, however setting the fixed direction, 5423 arrived. Taking these two factors into account, it would take approximately 16.4 days to obtain the number of sigmas shown in Fig. \ref{Fig_SigmaFraction} and Fig. \ref{Fig_SigmaWidth}.

\begin{figure}
\begin{center}
\includegraphics*[width=1\textwidth]{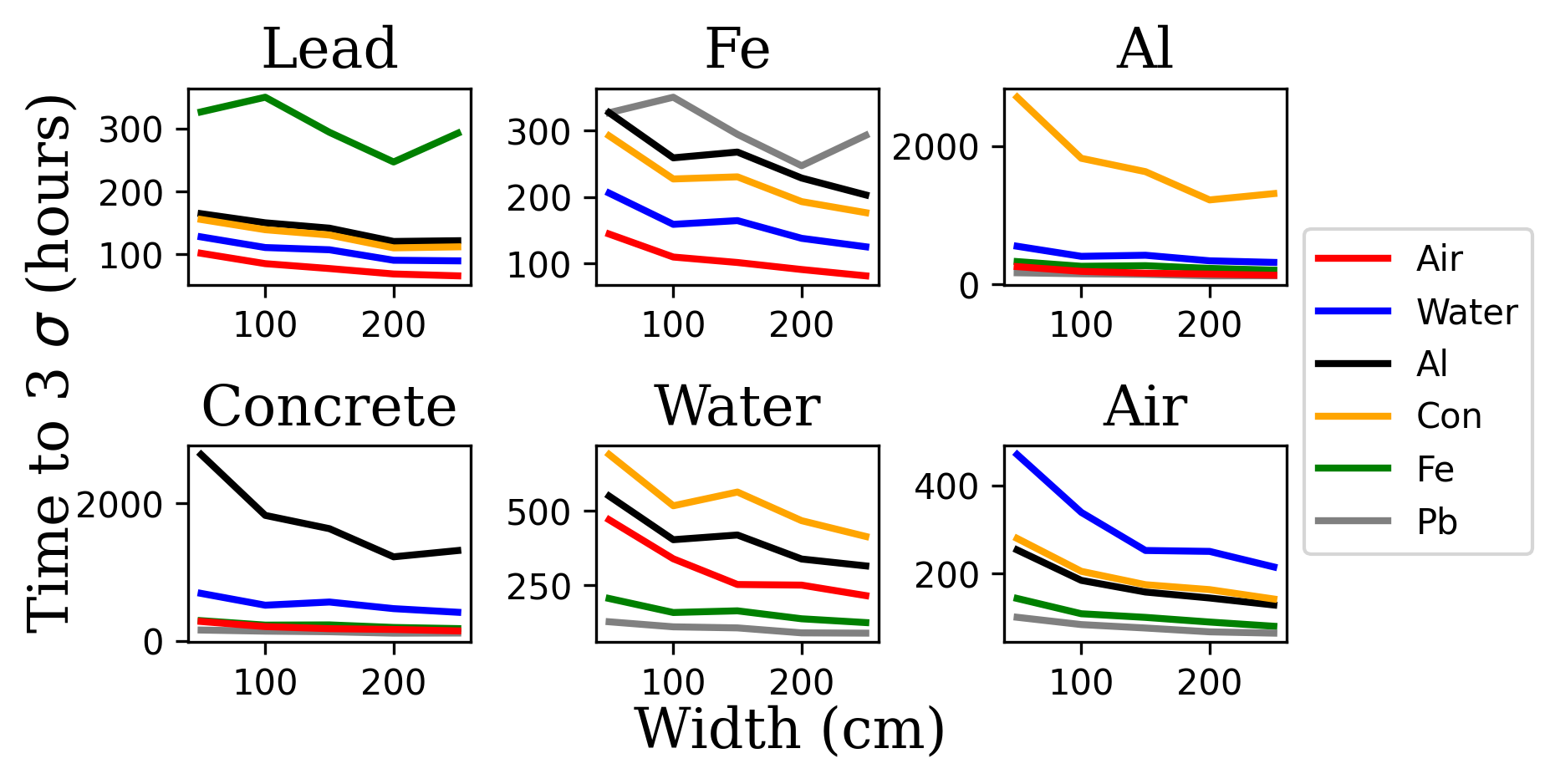}
\end{center}
  \caption{Time in hours to get 3 sigma deviations between the observed fraction of particles detected in the sub-detectors after the object and before it for different object materials.}
  \label{Fig_Fraction-Time}
\end{figure}

\begin{figure}
\begin{center}
\includegraphics*[width=1\textwidth]{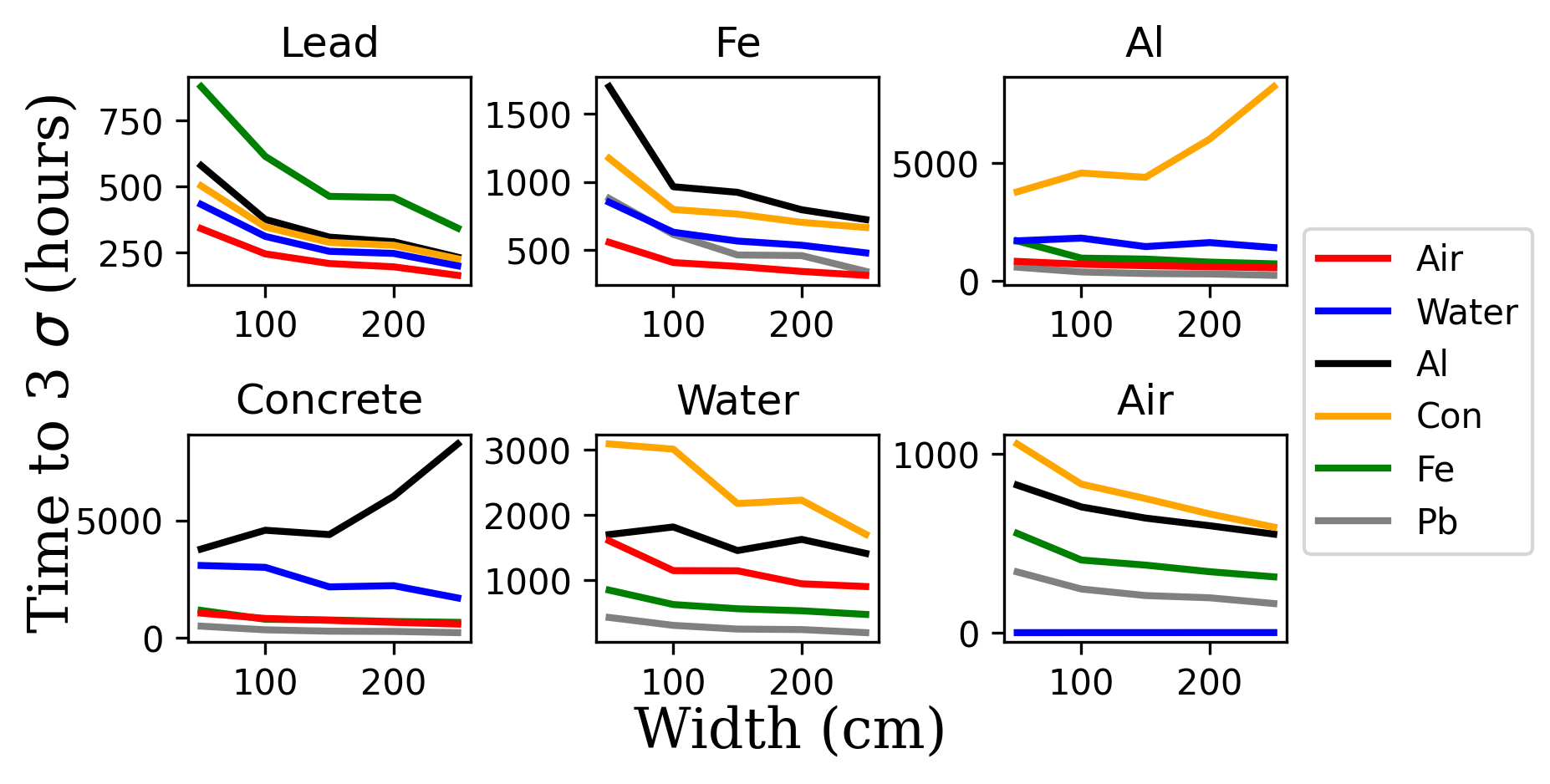}
\end{center}
  \caption{Time in hours to get 3 sigma deviations between widths of Gaussian fits of the histograms for scattering angles after passing through the object.}
  \label{Fig_Gauss-Time}
\end{figure}

The real number of hours to reach 3 $\sigma$ are shown in Fig. \ref{Fig_Fraction-Time} for the absorption method and in Fig. \ref{Fig_Gauss-Time} for scattering method. Setting as threshold less than 1 week for obtaining 3 $\sigma$ we get the following results for the absorption method:

   \begin{itemize}
     \item Between lead and air or water or aluminum or concrete.
     \item Between iron and air or water.
     \item Between air and aluminum or concrete.
     \end{itemize}

For the scattering method only between lead and air the acquisition time is less than 1 week for an object of 250 cm width.

\begin{figure}
\begin{center}
\includegraphics*[width=0.6\textwidth]{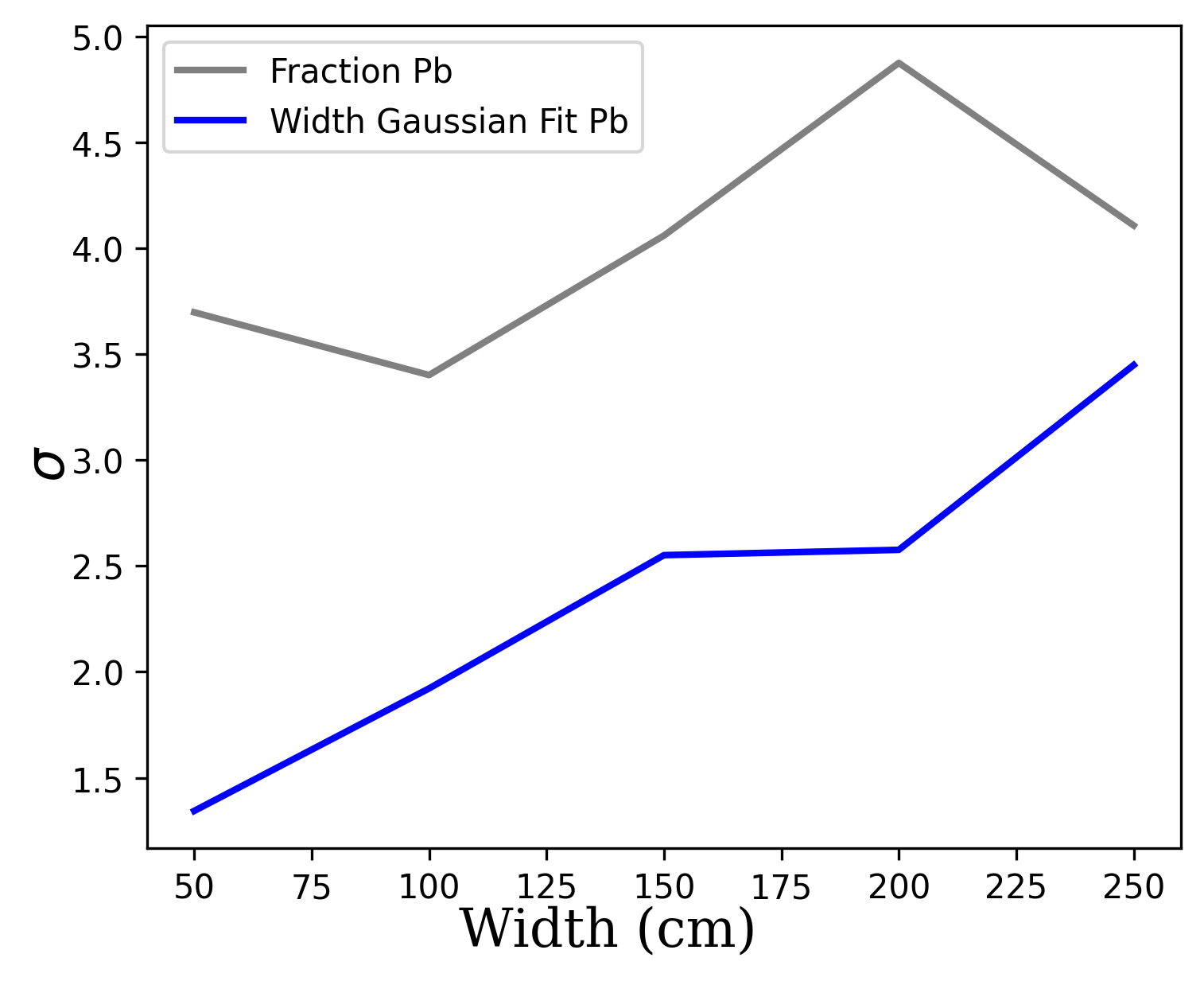}
\end{center}
  \caption{Number of sigmas comparing an object made of lead (Pb) and one made of iron (Fe) showing the two methods studied in this work. Grey line: absorption method and blue line: scattering method.}
  \label{2met}
\end{figure}

In order to compare both methods we show in Fig. \ref{2met} the number of sigmas versus the width of the object made of lead or iron. The absorption method has a larger difference between materials than the scattering method.

\section{Conclusions}
\label{sec:Conclu}
In this work, we have designed, simulated and analysed a detector easy to construct and transportable for muon tomography, an innovative technique in archaeology, mineralogy, geology, etc. for scanning mid to large objects. We used a flux of cosmic rays simulated in CORSIKA to examine how muons behave when passing through objects of several sizes made of different materials using GEANT4. 

Our detector is made of a plane array of 8 x 8 small unit sensors made of plastic scintillators and silicon photomultipliers enclosed in an aluminium case. Two of such plane arrays separated by 20 cm made a sub-detector. In order to analyze an object, one subdetector is placed before it and the second one after it. With this geometrical distribution the detector can measure the muon flux and the scattering angle of their trajectories after passing through an object. The sub-detector that is after the object will move around 90$^{\circ}$ to multiples angles to obtain more statistics. This rotation, thanks to the portability of this design, will increase the tested lines along the object, helping to find structures inside the object. However, in order to reduce costs when constructing the array of each plane, we could use instead of rectangular detectors, strips of plastic scintillators overlapping one under another. This detector has 1 $^{\circ}$ of angular resolution. Comparing with gas detectors, most of them have better angular resolution, nevertheless, moving gas detectors is more complicated.

The analyzed object had a varying size from 50 to 250 cm, which corresponds to a simplification of a larger and more complex object. Real objects can have holes or be made of different materials and have different geometries. These results already set a limit on which materials could be identified.

Our study demonstrated the efficiency of the fraction of detected particles using absorption methods and the width of the Gaussian fit for scattering method as reliable methods for material differentiation in muon tomography. Dense materials like lead can absorb a higher number of muons and muons scatter with a greater angle, showing as a result larger sigma values in their differentiation from other materials. 

The absorption method shows significant differences in material differentiation. Lead and iron are distinguishable from all other materials at different object widths. The scattering method shows lower values of sigma than the absorption method. Using this method, we obtain a substantial difference between an object made of different materials of all widths and an object of 250 cm width made of lead. We need less than 1 week with the absorption method for finding differences at 3 sigmas between lead or iron and other materials at all width of objects that we tested. For the scattering method at the same time and the same number of sigmas, we can find differences between lead and air. 

Although we can differentiate some materials, we recognize also the challenge of finding differences between aluminum and concrete, which exhibit only 0.9 sigmas with the absorption method and 0.1 with the scattering method at 250 cm of width of the object. This results shows the complexity of distinguishing between materials with close properties. 

Our study provides an understanding of material differentiation in the field of muon tomography. The absorption method is more robust and allows us better separation power between materials. However, this result can improve with more statistics. It is necessary to refine these techniques to reduce the observation time and expand the horizons of muon tomography applications.


\acknowledgments

J. R. appreciates the support from the Peruvian National Council for Science, Technology and Technological Innovation scholarship under Grant 23-2015-FONDECyT, thanks Dr. Hernan Asorey and Dr. Mauricio Suarez for their input on CORSIKA and GEANT4 and thanks again to Dr. Mauricio Suarez and Dr. Israel Martinez for reading the manuscript and their useful suggestions. J. B. thanks the Dirección de Fomento de la Investigación (DFI - PUCP) for funding under Grant No. DFI-2021-C-0020.


\bibliographystyle{JHEP}
\bibliography{biblio.bib}

\providecommand{\href}[2]{#2}\begingroup\raggedright\begin{thebibliography}{10}

\bibitem{George1956}
E.~George and G.~Shrikantia, \emph{Observations of the energy-spectrum of the
  cosmic radiation below ground},
  \href{https://doi.org/https://doi.org/10.1016/0029-5582(56)90129-8}{\emph{Nuclear
  Physics} {\bfseries 1} (1956) 54}.

\bibitem{Alvarez1970}
L.W.~Alvarez et~al., \emph{Search for hidden chambers in the pyramids},
  \href{https://doi.org/10.1126/science.167.3919.832}{\emph{Science} {\bfseries
  167} (1970) 832}.

\bibitem{Morishima}
K.~Morishima et~al., \emph{Discovery of a big void in {Khufu’s} pyramid by
  observation of cosmic-ray muons},
  \href{https://doi.org/10.1038/nature24647}{\emph{Nature} {\bfseries
  552(7685)} (2017) 386–390}.

\bibitem{Ambrosino_2014}
F.~Ambrosino et~al., \emph{The {MU-RAY} project: Detector technology and first
  data from {Mt. Vesuvius}},
  \href{https://doi.org/10.1088/1748-0221/9/02/c02029}{\emph{JINST} {\bfseries
  9} (2014) C02029}.

\bibitem{Pena-Rodriguez:2020}
J.~Peña-Rodríguez et~al., \emph{{Design and Construction of {MuTe}: a Hybrid
  Muon Telescope to study Colombian Volcanoes}},
  \href{https://doi.org/10.1088/1748-0221/15/09/P09006}{\emph{JINST} {\bfseries
  15} (2020) P09006} [\href{https://arxiv.org/abs/2004.09364}{{\ttfamily
  2004.09364}}].

\bibitem{Ola_2019}
L.~Oláh et~al., \emph{Muographic observation of density variations in the
  vicinity of {Minami-Dake} crater of {Sakurajima} volcano},
  \href{https://doi.org/10.20965/jdr.2019.p0701}{\emph{Journal of Disaster
  Research} {\bfseries 14} (2019) 701}.

\bibitem{Tanaka}
H.~Tanaka et~al., \emph{Development of the cosmic-ray muon detection system for
  probing internal-structure of a volcano},
  \href{https://doi.org/doi:10.1023/A:1020843100008}{\emph{Hyperfine
  Interactions} {\bfseries 138} (2001) 521–526}.

\bibitem{Woo2013}
W.J.~Jo et~al., \emph{Design of a muon tomography system with a plastic
  scintillator and wavelength-shifting fiber arrays},
  \href{https://doi.org/https://doi.org/10.1016/j.nima.2013.05.115}{\emph{Nucl.
  Instrum. Meth. A} {\bfseries 732} (568–572) 2013}.

\bibitem{Morris2014}
C.L.~Morris et~al., \emph{Horizontal cosmic ray muon radiography for imaging
  nuclear threats},
  \href{https://doi.org/10.1016/j.nimb.2014.03.017}{\emph{Nucl. Instrum. Meth.
  B} {\bfseries 330} (2014) 42}.

\bibitem{JungHyun}
J.~Bae et~al., \emph{Image reconstruction algorithm for momentum dependent muon
  scattering tomography},
  \href{https://doi.org/https://doi.org/10.1016/j.net.2023.12.009}{\emph{Nuclear
  Engineering and Technology} (2023) 1738}.

\bibitem{Wang_2024}
Z.~Wang, Y.~Wang, X.~Li, Y.~Zhao, Y.~Liang, Z.~Liang et~al., \emph{Electronics
  design for a muon imaging system using triangular plastic scintillators with
  wls fiber readouts},
  \href{https://doi.org/10.1088/1748-0221/19/02/P02033}{\emph{JINST} {\bfseries
  19} (2024) P02033}.

\bibitem{Yu:2024spj}
X.~Yu et~al., \emph{{A proposed PKU-Muon experiment for muon tomography and
  dark matter search}},  \href{https://arxiv.org/abs/2402.13483}{{\ttfamily
  2402.13483}}.

\bibitem{Chaiwongkhot}
K.~Chaiwongkhot et~al., \emph{{3D} cosmic-ray muon tomography using portable
  muography detector},
  \href{https://doi.org/10.1088/1748-0221/17/01/P01009}{\emph{JINST} {\bfseries
  17} (2022) }.

\bibitem{Bajou}
R.~Bajou et~al., \emph{High-resolution structural imaging of volcanoes using
  improved muon tracking},
  \href{https://doi.org/https://doi.org/10.1093/gji/ggad269}{\emph{Geophys. J.
  Int.} {\bfseries 235} (2023) 1138–1149}.

\bibitem{Barnes}
S.~Barnes et~al., \emph{Cosmic-ray tomography for border security},
  \href{https://doi.org/https://doi.org/10.3390/instruments7010013}{\emph{Instruments}
  {\bfseries 13} (2023) }.

\bibitem{HAL_Muons}
J.L.~Autran et~al., \emph{Characterization of atmospheric muons at sea level
  using a cosmic ray telescope},
  \href{https://doi.org/10.1016/j.nima.2018.06.038}{\emph{Nucl. Instrum. Meth.
  A} {\bfseries 903} (2018) 77}.

\bibitem{Lorenzo_2020}
L.~Bonechi et~al., \emph{Atmospheric muons as an imaging tool},
  \href{https://doi.org/10.1016/j.revip.2020.100038}{\emph{Reviews in Physics}
  {\bfseries 5} (2020) }.

\bibitem{GroomPDG2}
D.E.~Groom, N.~Mokhov and S.~Striganov, \emph{Muon stopping power and range
  tables 10 {MeV - 100 TeV}}, {\emph{Atomic Data and Nuclear Data Tables}
  {\bfseries 76 (2)} (2001) }.

\bibitem{Fredrick2016}
H.F.~Schreiner, \emph{Methods and Simulations of Muon Tomography and
  Reconstruction}, Ph.D. thesis, University of Texas at Austin, 2016.

\bibitem{Lynch_1991}
G.R.~Lynch et~al., \emph{Approximations to multiple {Coulomb} scattering},
  \href{https://doi.org/10.1016/0168-583x(91)95671-y}{\emph{Nucl. Instrum.
  Meth. B} {\bfseries 58} (1991) 6–10}.

\bibitem{Kaiser2019}
R.~Kaise, \emph{{Muography: Overview and Future Directions}},
  \href{https://doi.org/http://doi.org/10.1098/rsta.2018.0049}{\emph{Philos.
  Trans. R. Soc. A} {\bfseries 377} (2019) }
  [\href{https://arxiv.org/abs/1905.12311}{{\ttfamily 1905.12311}}].

\bibitem{Axani:2018qzs}
S.N.~Axani et~al., \emph{{The CosmicWatch Desktop Muon Detector: a
  self-contained, pocket sized particle detector}},
  \href{https://doi.org/10.1088/1748-0221/13/03/P03019}{\emph{JINST} {\bfseries
  13} (2018) P03019} [\href{https://arxiv.org/abs/1801.03029}{{\ttfamily
  1801.03029}}].

\bibitem{Axani2017}
S.N.~Axani et~al., \emph{The desktop muon detector: A simple, physics-motivated
  machine- and electronics-shop project for university students},
  \href{https://doi.org/10.5194/gi-1-185-2012}{\emph{Am. J. Phys.} {\bfseries
  85} (2017) 948}.

\bibitem{Hachaj2023}
T.~Hachaj and M.~Piekarczyk, \emph{The practice of detecting potential cosmic
  rays using cmos cameras: Hardware and algorithms},
  \href{https://doi.org/https://doi.org/10.3390/s23104858}{\emph{Sensors}
  {\bfseries 23(10)} (2023) 4858}.

\bibitem{Poulson}
D.~Poulson et~al., \emph{Cosmic ray muon computed tomography of spent nuclear
  fuel in dry storage casks.},
  \href{https://doi.org/doi:10.1016/j.nima.2016.10.040}{\emph{Nucl. Instrum.
  Meth. A} {\bfseries 842} (2017) 48–53}.

\bibitem{Clarkson}
A.~Clarkson et~al., \emph{The design and performance of a scintillating-fibre
  tracker for the cosmic-ray muon tomography of legacy nuclear waste
  containers.},
  \href{https://doi.org/doi:10.1016/j.nima.2014.01.062}{\emph{Nucl. Instrum.
  Meth. A} {\bfseries 745} (2014) 138–149}.

\bibitem{Heck_CORSIKA}
D.~Heck et~al., \emph{{CORSIKA}: {A} {M}onte {C}arlo {C}ode to {S}imulate
  {E}xtensive {A}ir {S}howers}, vol.~6019, Karlsruhe Institute of Technology
  (1998).

\bibitem{MF_NOAA}
{NOAA}, \emph{National centers for environmental information of {NOAA}},  2020.

\bibitem{Asorey_Data_Access}
{H. Asorey}, \emph{The {LAGO} {CrkTools} suite},  2019.

\bibitem{Rigiditty2021}
Izmiran, \emph{Cutoff rigidity calculator},  2021.

\bibitem{Agostinelli2003}
{\scshape GEANT4} collaboration, \emph{{{GEANT4--a simulation toolkit}}},
  \href{https://doi.org/10.1016/S0168-9002(03)01368-8}{\emph{Nucl. Instrum.
  Meth. A} {\bfseries 506} (2003) 250}.

\bibitem{ALTAMEEMI2019281}
N.I.~Rasha et~al., \emph{Determination of muon absorption coefficients in heavy
  metal elements},
  \href{https://doi.org/https://doi.org/10.1080/16878507.2019.1652965}{\emph{J.
  Radiat. Res. Appl. Sci.} {\bfseries 12} (2019) 281}.

\end{thebibliography}\endgroup

\end{document}